\documentclass[aps, reprint, amsmath, amssymb, superscriptaddress]{revtex4-1}

\usepackage{amsmath}
\usepackage{amsfonts}
\usepackage{amssymb}
\usepackage{amsthm}
\usepackage{fancyhdr}
\usepackage{enumerate}
\usepackage[top=1in, bottom=1in, left=1in, right=1in]{geometry}
\usepackage{graphicx}
\usepackage[labelfont=bf]{caption}

\usepackage{color}

\usepackage{caption}
\usepackage{subcaption}

\usepackage{verbatim}

\usepackage{wasysym}
\usepackage{hyperref}

\usepackage{wrapfig}

\newcommand{\Exp}[1]{\langle #1 \rangle}
\newcommand{\K}{\mathcal{K}}
\newcommand{\F}[1]{{}_1\mathcal{F}_1\left(#1\right)}
\newcommand{\FReg}[1]{{}_1\mathcal{F}^{R}_1\left(#1\right)}
\newcommand{\Fz}{{}_1\mathcal{F}_1}
\newcommand{\FRegz}{{}_1\mathcal{F}^{R}_1}
\newcommand{\yb}{ \bar y}

\begin{document}

\title{Population Extinction on a Random Fitness Seascape}
\author{Bertrand Ottino-L\"{o}ffler and Mehran Kardar}
\affiliation{Department of Physics, Massachusetts Institute of Technology, Cambridge,
Massachusetts 02139, USA}
\date{\today}

\begin{abstract}
Models of population growth and extinction are an increasingly popular subject of study. 
However, consequences of stochasticity and noise in shaping distributions and outcomes are not sufficiently explored.
Here we consider a distributed population with logistic growth at each location, subject to 
``seascape'' noise, wherein the population's fitness randomly varies with {\it  location and time}. 
Despite its simplicity, the model actually incorporates variants of directed percolation, and directed
polymers in random media, within a mean-field perspective.
Probability distributions of the population can be computed self-consistently;
and the extinction transition is shown to exhibit novel critical behavior with exponents
dependent on the ratio of the strengths of migration and noise amplitudes.
The results are compared and contrasted with the more conventional choice of demographic noise
due to stochastic changes in reproduction.
\end{abstract}

\maketitle

\section{Introduction}
The growth of spatially distributed populations is of importance to ecology, evolutionary biology, epidemiology, 
among many other fields. One of the simplest models of a species or pathogen propagating in 
space is the \emph{ Fisher equation}~\cite{fisher1937wave, aronson1978multidimensional, gourley2000travelling, kwapisz2000uniqueness, hallatschek2009fisher} given by
\begin{equation}\label{Eq:Fisher}
\frac{dy(x,t)}{dt} = \mu y - a y^2 + D \nabla^2 y~.
\end{equation}
At each spatial coordinate $x$, the population size $y(x,t)$  initially increases exponentially in time $t$,
at rate determined by the {\it fitness parameter} $\mu>0$. This (logistic type) growth is eventually
slowed down by the nonlinear term, describing resource limitations, with the population asymptoting to $\mu/a$ (the capacity). 
Alternatively, for $\mu<0$, the population decays to zero.
The diffusion term $D$ captures migration of the population to nearby points in space.

Such deterministic evolution is only valid for populations of infinite size for which fluctuations
can be ignored. In particular, near extinction points where the population hits zero 
(such as when a species suffers ecological collapse or when a pathogen is without available hosts),
the deterministic description does not capture the stochasticity of small number fluctuations.
Transitioning from a model with a finite, discrete population into a continuum formulation 
results in a noise of amplitude proportional to the square root of the number of individuals.
Including this so-called \emph{demographic noise}~\cite{durrett1994importance,butler2009robust,traulsen2012stochastic,constable2016demographic,weissmann2018simulation}
leads to the stochastic Fisher equation 
\begin{equation} \label{basic_demog_space}
dy = \left(\mu y - ay^2 + D \nabla^2 y\right)dt + \sigma_d \sqrt{y} dW~,
\end{equation}
with $\sigma_d$ as the noise amplitude, and $dW = dW(x,t)$ being a standard Wiener process. 
Models of this type have been studied in a number of contexts outside of population growth, 
thanks to  connections to directed percolation~\cite{cardy1980directed, tauber1998multicritical, janssen2005field,  peschanski2009traveling, sipos2011directed, korolev2010genetic, horowitz2019bacterial}
(with onset of laminar turbulence for 
$\mu = a = 0$~\cite{sipos2011directed} as a recent application).
Indeed, extinction transitions are expected to generically belong to the
directed percolation universality class.~\cite{Henkel2008DP,Tauber2014DP}

The main focus of this paper is on a different form of noise. 
The implicit assumption of the Fisher equation is that the fitness  $\mu$ is an intrinsic
property, uniform in time and the same across all locations.
For most populations, however, the growth rate is strongly influenced by external factors,
such nutrient level, light, etc. at different times and places.
Generalizing from a static but position dependent {\it fitness landscape} $\mu(x)$,
a time-varying fitness $\mu(x,t)$ is referred to as a  \emph{seascape}~\cite{mustonen2009fitness, mustonen2010fitness, iram2019controlling, agarwala2019adaptive}.
Population growth on a random seascape is thus governed by
\begin{equation} \label{basic_param_space}
dy = \left(\mu y - ay^2 + D \nabla^2 y\right)dt + \sigma y dW,
\end{equation}
with $\sigma^2$ being the variance of the  fitness noise, and $dW = dW(x,t)$ a Wiener process as before. 
Note that for $a=0$, the linear equation describes the evolving weight $y(x,t)$ of a directed polymer
in a quenched random energy landscape ($\sigma dW(x,t)$); with $\ln y(x,t)$ satisfying
the Kardar-Parisi-Zhang equation, notable in the study of surface growth 
problems~\cite{kardar1986dynamic, kardar1987scaling, derrida1990directed, chu2016probability, borodin2014free, medina1989burgers, quastel2015one, sasamoto2010one, halpin2013extremal}. 

While much studied, the problem of directed polymers in random media is largely unsolved in
other than one dimension. Such limitation will necessarily extend to characterizing the effect of a random seascape
on a population in two or more dimensions. 
We thus resort to replacing near-neighbor migrations with jumps of arbitrary length.
This ``mean-field" limit is reasonable for a well mixed bacterial population, 
or for a pathogen infecting an active city.
An alternative description is to replace continuous space with a series of nodes indexed by
$k=1,2,\cdots,N$. If population $y_k$ of each node can jump to any other node in the network
at rate $D/N$, then (for $N\gg 1$), evolution in a random seascape is described by 
\begin{equation} \label{basic_param}
dy_k = \{ \mu y_k - a{y_k}^2 + D (\bar y - y_k) \}dt + \sigma y_k dW_k~,
\end{equation}
and similarly for demographic noise we have
\begin{equation} \label{basic_demog}
dy_k = \{\mu y_k - a{y_k}^2 + D (\bar y - y_k) \}dt + \sigma_d \sqrt{y_k} dW_k~,
\end{equation}
where $\bar y = \frac{1}{N} \sum_k y_k$ is the spatial average, with $dW_k = dW(k,t)$ being spatiotemporal noise. 
In a sense, the diffusion coefficient $D$ now behaves like a mean-field coupling strength. 
The problem can be solved exactly in this case, as for $N\to\infty$,  the probability distribution for $\bar y$
converges to a $\delta$-function, reducing the above equations to those of independent variables whose distributions
can be obtained by standard methods. The important population average  $\bar y$ is then computed
self-consistently.

We carry our this program below initially for seascape noise, in the three  cases of:
neural evolution  ($\mu=a=0$) in Sec.~\ref{par_0_sec}, decaying population ($\mu < 0$) in Sec.~\ref{par_neg_sec},
and saturating growth ($\mu > 0$, $a>0$) in Sec.~\ref{par_pos_sec}.   
For comparison, the cases of demographic noise, as well as mixed demographic and seascape noise, are
studied Sec.~\ref{dem_sec}.   

Despite the commonality in their deterministic parts, the two types of stochastic noise relate  to two 
quite distinct universality classes; that of directed polymers in random media for Eq.~\eqref{basic_param},
versus directed percolation for Eq.~\eqref{basic_demog}.
As such, we might expect them to exhibit different behaviors in the all-important extinction regime. 
This is indeed the case: 
Near the extinction threshold, we show that seascape noise leads to a broadly distributed population, characterized
by a tail falling off as a power-law. The power-law exponent varies continuously with $\sigma^2/D$ such
that, when the noise is large, fluctuations in the population much exceed the mean.
The extinction transition no longer belongs to the (``mean-field") directed percolation universality class.
However, as  we show in Sec.~\ref{dem_sec}, demographic noise restores the directed percolation universality class. 

\section{Neutral population for $\mathbf{\mu = a = 0}$} \label{par_0_sec}

\begin{figure}[t]
\centering
\includegraphics[width = 0.5\textwidth]{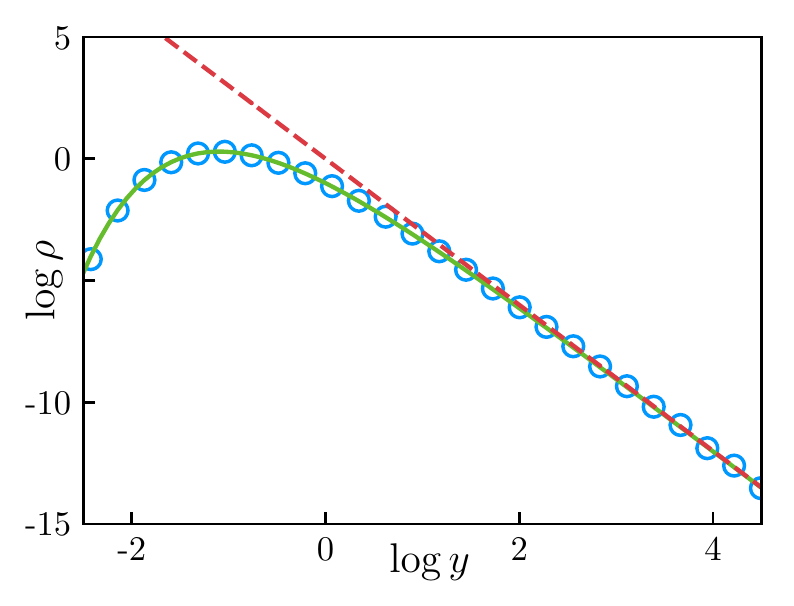}
\caption{Simulated versus predicted probability distribution in the
steady-state of Eq.~\eqref{dyn_MF}. Circles show simulation results from $5 \times 10^6$ runs of Eq.~\eqref{dyn_MF_inf}, the solid line shows the prediction of Eq.~\eqref{MF_pdf}, 
while the dashed line emphasizes the  large-$y$ behavior from Eq.~\eqref{MF_tail}. Simulation was run using stochastic Runge-Kutta, with $y_0 = 1$, $D = 0.5$, $\sigma = 1.0$, and $\mu = a = 0$. The simulation ran for 25 time units, with a time-step of $0.01$.}
\label{MF_density_plot}
\end{figure}

\begin{figure}[t]
\centering
\includegraphics[width = 0.5\textwidth]{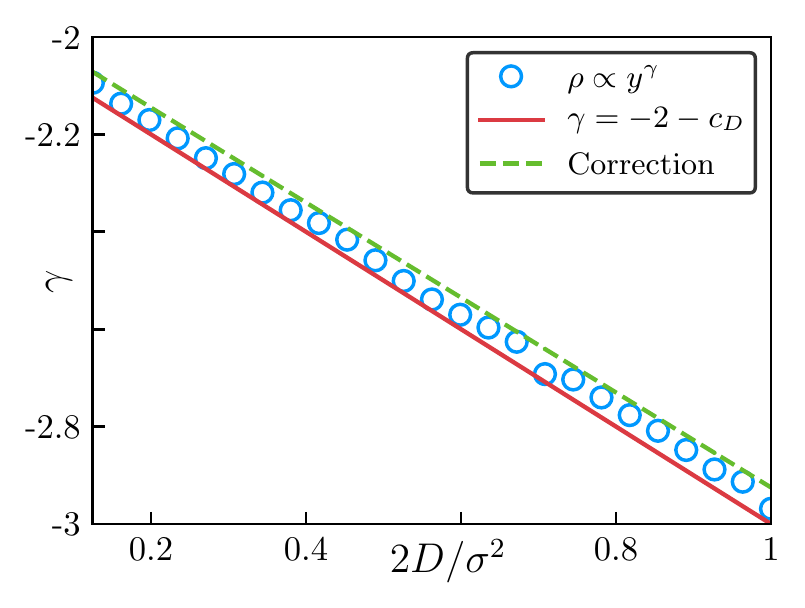}
\caption{Simulated versus predicted power-law decays of the tail of the probability distribution in the
steady-state of Eq.~\eqref{dyn_MF}.
Circles show numerically obtained decay exponents from simulating $5 \times 10^6$ replications of Eq.~\eqref{dyn_MF_inf}; the solid line depicts the  predicted decay exponent at large $y$ from Eq.~\eqref{MF_tail}; 
and the dashed line shows the maximum error arising from measuring the exponent at finite $y$, 
as predicted via Eq.~\eqref{MF_pdf}. The simulation was run using stochastic Runge-Kutta, with 
$y_0 = 1$, $D = 0.5$, $\sigma = 1.0$, and $\mu = a = 0$. The time-step of integration was $0.01$, with a total integration time of $25$ units. Regression to identify the decay rate of the PDFs began at $y^* \approx 13.34$, 
and the final exponent was averaged over the final 5 time units with samples every 0.05 time units.}
\label{MF_tail_plot}
\end{figure}

Many subtleties of seascape noise can be gleamed by considering Eq.~\eqref{basic_param} in the 
`neutral' case, given by
\begin{equation}\label{dyn_MF}
dy_k =D \left(\bar y - y_k \right)dt + \sigma y_k dW_k~.
\end{equation}
The above equation can be interpreted as describing a form of synchronization: 
each $y_k$ is being pulled towards the spatial mean $\bar y$ with strength $D$, 
but they are forced apart by the noise  $\sigma y_k dW_k$. 
The interplay between the two tendencies manifests in the population variance, 
with one limiting form being a ``synchronized'' population described by a tight distribution, and the other being 
an ``incoherent'' population that is broadly distributed~\cite{strogatz2004sync, pikovsky2003synchronization}.

While it would be interesting to analyze the case of a finite $N$, 
 computations simplify in the thermodynamic limit of $N\to\infty$. 
If starting with a uniform initial condition, say $y_k(t= 0) \equiv y_0$ for all $k$, 
then the evolving probability distributions of all individual members  will be identical. 
Thus, if there exists some long-time stationary distribution with finite mean, 
then this mean will be the same for each member. So by simply applying the law of large numbers and taking expectations, we obtain $\bar y = \Exp{y} = y_0$.
The second equality follows from the conservation of average population in the neutral limit,
remaining at $Ny_0$ at all times.

Therefore, the large-$N$ dynamics are captured by the single stochastic differential equation
\begin{equation}\label{dyn_MF_inf}
dy = D(y_0 - y)dt + \sigma y dW~,
\end{equation}
independent of the spatial index $k$. 
From here, we can construct a Fokker-Plank equation for the evolution of the
probability density function $\rho(y,t)$, namely 
\begin{equation} \label{MF_FK}
\partial_t \rho = -\partial_y \left[ D(y_0 - y)\rho  - \frac{\sigma^2}{2} {\partial_y}\left(y^2 \rho\right)\right]~.
\end{equation}
A stationary distribution ($\partial_t \rho \equiv 0$) is obtained by setting the probability current
in the above square bracket to zero, leading to
\begin{equation} \label{MF_FK_dyn}
0 = y^2 \rho' + (c_D +2) y \rho - c_D y_0 \rho~,
\end{equation}
where we define $c_D = 2D/\sigma^2$. 
This is a solvable ordinary differential equation, with a properly normalized solution given by
\begin{equation}\label{MF_pdf}
\rho(y) = \frac{\left(c_D y_0\right)^{c_D+1}}{\Gamma(c_D+1)} y^{-2-c_D} e^{-c_D y_0 /y}~,
\end{equation}
which is a scaled inverse Chi-squared distribution with $\nu = 2(c_D+1)$ and $\tau = y_0 c_D/(1+c_D)$. This prediction agrees with simulated infinite-$N$ dynamics depicted in Fig.~\ref{MF_density_plot}.

An important characteristic of this solution is the behavior in the tail, where for large populations $y$, the density 
behaves as 
\begin{equation}\label{MF_tail}
\rho \propto y^{-2-c_D}~;
\end{equation} 
confirmed in Fig.~\ref{MF_tail_plot}. 
Even though the mean of the distribution stays at the initial value of $y_0$, 
the seascape fluctuations lead to a power-law tail which results in
a variance (and  higher moments) which may either be finite or infinite! 
Specifically, we have 
\begin{equation} \label{MF_ratio}
\frac{\Exp{y^2}}{\Exp{y}^2} = \frac{2D}{2D - \sigma^2}~,
\end{equation}
such that the variance becomes infinite when $c_D$ crosses 1, and the population distribution 
becomes broad compared to the mean. 
Parenthetically, we note that such a ``transition" does not occur in case of demographic noise
as described later in the text.

\section{Decaying population  for $\mathbf{\mu < 0}$} \label{par_neg_sec}

\begin{figure}[h]
\centering
\includegraphics[width = 0.5\textwidth]{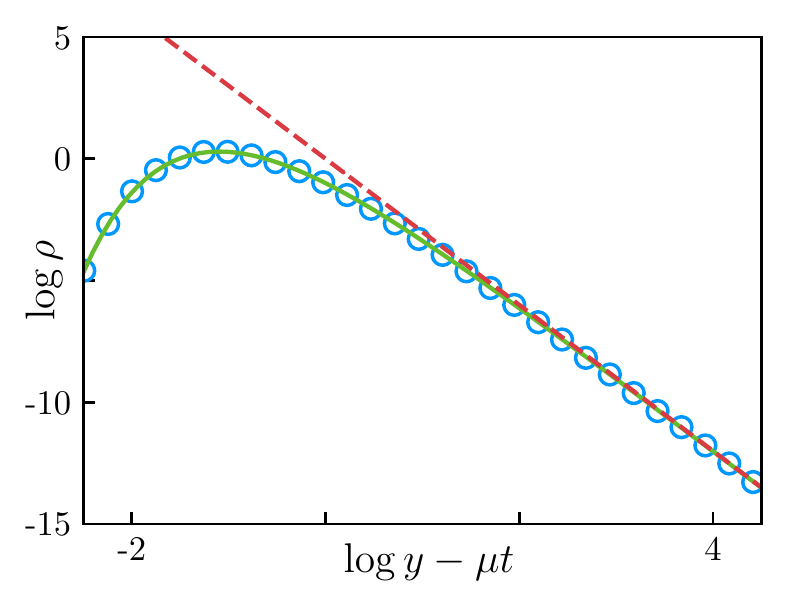}
\caption{Simulated versus predicted probability densities for the decaying population;
note the time-dependent horizontal axis. 
Circles show simulation results from $5 \times 10^6$ runs of Eq.~\eqref{basic_param};
the solid line depicts the exact steady-state solution of Eq.~\eqref{MF_pdf}; 
and the dashed line shows the predicted large-$y$ behavior from Eq.~\eqref{MF_tail}. 
Simulation was run using stochastic Runge-Kutta, with $y_0 = 1$, $D = 0.5$, 
$\sigma = 1.0$, $\mu = -0.1$, and $a = 1$. 
The time-step of integration was $0.01$, with a total integration time of $100$ units.}
\label{negative_density_plot}
\end{figure}

Now consider Eq.~\eqref{basic_param} in the presence of negative fitness $\mu$. 
The nonlinear term is now asymptotically irrelevant as can be seen by the transformation to
$z = ye^{-\mu t}$, such that the dynamics in this `moving frame' becomes
\begin{equation} \label{Neg_dyn_trans1}
dz = \left[ -a z^2 e^{\mu t} + D\left( \bar z -z\right) \right]dt + \sigma z dW~.
\end{equation}
Since $\mu <0$, the nonlinear term drops out at large times, leaving us with 
\begin{equation} \label{Neg_dyn_trans2}
dz = D\left( \bar z -z\right)dt + \sigma z dW~,
\end{equation}
which is simply the exact same dynamics as for $\mu = 0$ in Eq.~\eqref{dyn_MF}. 
Therefore, the same long-time results hold  in these transformed coordinates. 
This  includes the exact distribution predicted in Eq.~\eqref{MF_pdf} as well as the moment ratio
in Eq.~\eqref{MF_ratio}, as confirmed by the simulations depicted in Fig.~\ref{negative_density_plot}. 
Therefore, while the mean population size decays exponentially, the ratio
\begin{equation} \label{Neg_ratio}
\frac{\Exp{y^2}}{\Exp{y}^2} = \frac{2D}{2D - \sigma^2},
\end{equation}
indicates that the mean ceases to be a good measure of the distribution for
$2D/\sigma^2 < 1$.

\section{ Growing population  for $\mathbf{\mu > 0}$} \label{par_pos_sec}

The most intricate case corresponds to when  $\mu$, $a$, $D$, and $\sigma$ are all strictly positive:
\begin{equation}\label{Pos_dyn}
dy_k = \{ \mu y_k - a {y_k}^2 + D \left(\bar y - y_k \right)\}dt + \sigma y_k dW_k~.
\end{equation}
The intricacy arises as the noise modifies the mean value of $y$ from the bare value of $\mu/a$.
In reality, the dynamics of the mean depend on the value of every higher moment. 
More intuitively, note that the restoring force of the logistic growth term is asymmetric, 
in that it ``punishes'' overly large values of $y$ harder than small ones. 
Under the effect of noise, this biases $y$ to smaller values, and we should expect the true value of 
$\Exp y$ to  be smaller than $\mu/a$. 

To properly analyze this case, we need to construct a self-consistent solution in the long-time, large--$N$ limit. 
Assuming a long-time stationary distribution exists, and given that every node has the same initial condition $y_k(t=0) = y_0$, then the uniformity of their stochastic dynamics indicates that all nodes arrive
to the same distribution $\rho$, for all $k$, i.e. $\bar y = \int y \rho(y) dy$.
The stationary distribution $\rho(y)$ is obtained by examining the  Fokker-Plank equation
associated with the stochastic differential equation
\begin{equation}\label{Pos_inf_dyn}
dy = [ \mu y - a {y}^2 + D \left( \bar y  - y \right) ]dt + \sigma y dW~,
\end{equation}
given by
\begin{equation} \label{MF_FKb}
\partial_t \rho = -\partial_y \left[ \left( \mu y - a {y}^2+D(\bar y - y)\right) \rho  - \frac{\sigma^2}{2} {\partial_y}\left(y^2 \rho\right)\right]~.
\end{equation}
Setting the probability current to zero leads to the ordinary differential equation
\begin{equation}\label{Pos_FK_dyn}
0 = y^2 \rho' + ( 2 + c_D - c_\mu) y \rho - c_D y_0 \rho + c_a \rho y^2~,
\end{equation}
where $c_D = 2D/\sigma^2$, $c_\mu = 2\mu/\sigma^2$, and $c_a = 2a/\sigma^2$. 
This admits an unnormalized solution of the form 
\begin{equation}\label{Pos_pdf}
\hat \rho(y) = e^{-c_D \bar y /y} e^{-c_a y} y^{-2 - c_D +c_\mu}~.
\end{equation}
Note that finite $\bar y$ and $a$ cut off power-law tails of the distribution for small and large $y$,
respectively, so that all moments of the distribution are now finite.

\begin{figure}[t]
\centering
\includegraphics[width = 0.5\textwidth]{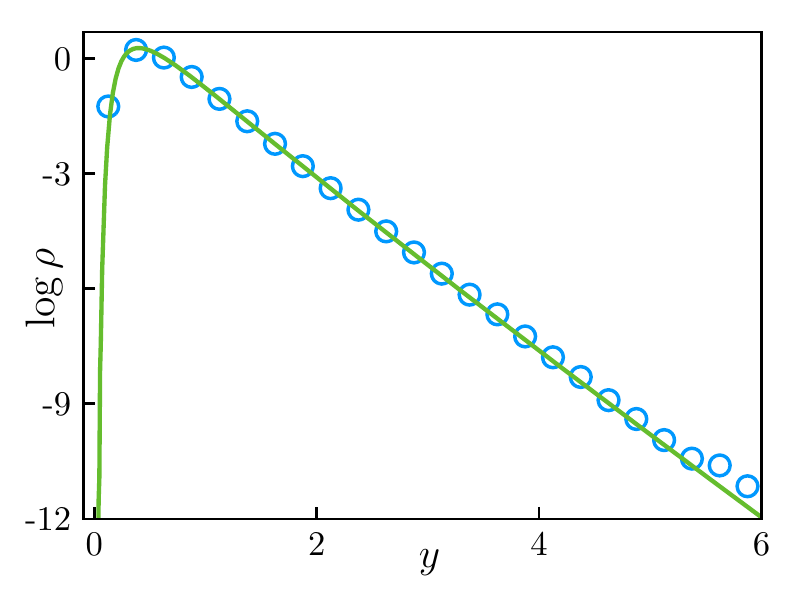}
\caption{Simulated versus predicted probability densities for Eq.~~\eqref{Pos_inf_dyn}. 
 Circles show simulation results from $5 \times 10^6$ runs; 
 the solid line depicts the exact steady--state solution of Eq.~\eqref{Pos_pdf}.
 Simulation was run using stochastic Runge-Kutta, with $D = 0.5$, $\sigma= 1$, $\mu = 1$ and $a = 1$. 
 The value of $\bar y$ was numerically decided in order to obey the self-consistency condition
 in Eq.~\eqref{Pos_Mean}. The time-step of integration was $0.01$, with a total integration time of $25$ units.
}
\label{positive_density_plot}
\end{figure}

To convert Eq.~\eqref{Pos_pdf} to a proper PDF, we need the normalization factor
\begin{alignat}{1}
Z &= \int_{0}^{\infty} \hat \rho(y) dy \notag \\
&= 2 \left( \frac{c_a}{c_D \bar y} \right)^{\frac{1 + c_D - c_\mu}{2}} \mathcal{K}_{1+c_D-c_\mu} \left(2 \sqrt{c_D c_a \bar y} \right)~, \label{Pos_norm}
\end{alignat}
where $\mathcal{K}_\gamma$ is a modified Bessel function of the second kind. 
The mean of the distribution is then given by
\begin{alignat}{1}
\Exp y Z &= \int_{0}^{\infty} y \hat \rho(y) dy \notag \\
&= 2 \left( \frac{c_a}{c_D \bar y} \right)^{\frac{ c_D - c_\mu}{2}} \mathcal{K}_{c_D-c_\mu} \left(2 \sqrt{c_D c_a \bar y} \right)~, \label{Pos_mean}
\end{alignat}
leading to the self-consistency equation, 
\begin{equation}\label{Pos_Mean}
\bar y = \frac{c_D}{c_a} \left( \frac{\mathcal{K}_\beta(2 x)}{\mathcal{K}_{\beta+1}(2 x)} \right)^2~,
\end{equation}
where we define $\beta = c_D - c_\mu$ and $x = \sqrt{c_D c_a \bar y}$ for the sake of convenience. 
The above equation can be solved numerically; as depicted in Fig.~\ref{positive_density_plot},
there is excellent agreement between the thus analytically constructed PDF and the result from
of numerical simulation.

\begin{figure}[t]
\centering
\includegraphics[width = 0.5\textwidth]{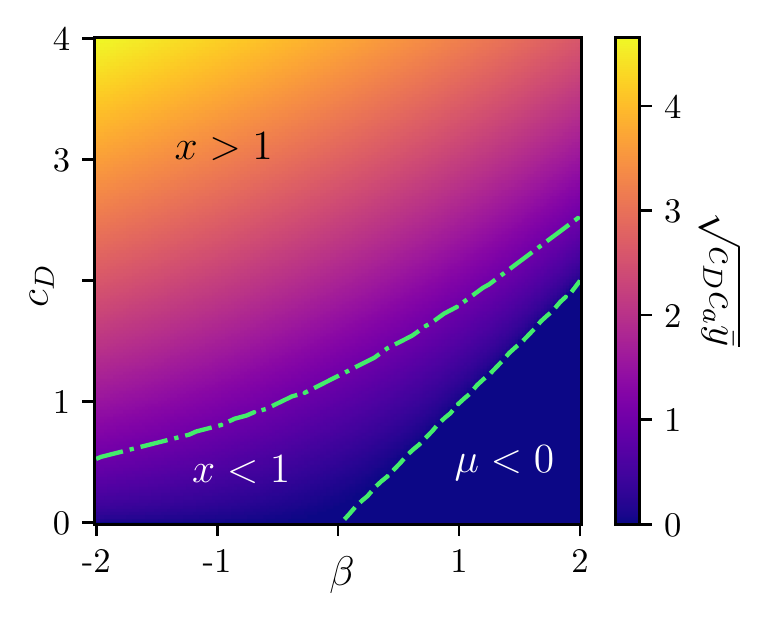}
\caption{`Heat map' depicting the dependence of the self-consistent mean  $\bar y$ from Eq.~\eqref{Pos_Mean} 
on the parameters $\beta$ and $c_D$.
The dashed line marks the boundary of the region where $\mu <0$,  and the population goes extinct.
The dash-dotted line shows where the quantity $x = \sqrt{c_Dc_a\bar y}$ is less than 1, 
denoting the region where a perturbative analysis may be appropriate. 
(The parameters $a= 1$ and $\sigma = 1$ were used for this plot.)}
\label{param_heat_plot}
\end{figure}

\subsection{Asymptotic behavior}
Going beyond the numerical solution of the self-consistency condition, we would like to gain analytic insight 
into the behavior of the population, particularly close to the extinction line.
To do so, it is convenient to rescast Eq.~\eqref{Pos_Mean} as
\begin{equation}\label{Pos_Consistent}
x = c_D \frac{\mathcal{K}_\beta(2x)}{\mathcal{K}_{\beta+1}(2x)}.
\end{equation}
The solution now explicitly depends on only two parameters, as $x = f(\beta, c_D)$, 
and can be plotted as in Fig.~\ref{param_heat_plot}. 
Notably, we focus on the region where $x$ is small, 
hoping to construct perturbative forms for small $x\propto\sqrt{\bar y}$.

As depicted in Fig.~\ref{param_heat_plot}, the small $x$ region spans two segments;
one corresponding to extinction as $\mu\to0$ along the line $c_D=\beta>0$,
and another corresponding to $c_D\to 0$ for $\beta<0$ corresponding to $\bar y\to 0$ as $a\to\infty$.
Only the former is of interest, and for which we need the asymptotic behavior of
the right hand side of Eq.~\eqref{Pos_Consistent} as $x\to 0$.
Not surprisingly, given the power-law in Eq.~\eqref{Pos_pdf}, the expansion of  
$\mathcal{K}_\beta/\mathcal{K}_{\beta+1}$ for $\beta > 0$ is non-analytic,
and given by
\begin{equation} \label{bess_ratio}
\frac{\mathcal{K}_\beta(2x)}{\mathcal{K}_{\beta+1}(2x)} \simeq x \left(\frac{1}{\beta} + \frac{x^2}{\beta^2(1-\beta)} + \frac{\Gamma(-\beta)}{\Gamma(\beta+1)} x^{2\beta} \right)~.
\end{equation}
The leading term in an expansion of Eq.~\eqref{Pos_Consistent} is thus $c_D/\beta$, indicating
that a non-zero solution exists for $c_D/\beta>1$. However, the next order term can be either
$x^3$ or $x^{1+2\beta}$ depending on $\beta$, setting up two distinct singular behaviors:

\begin{itemize}
\item{} For $\beta>1$ balancing  of the cubic and linear terms leads to 
$x^2\simeq \beta(1-\beta)c_\mu/c_D$. As $\mu\to 0$, the mean population size thus vanishes as
\begin{equation}\label{pos_mean_eq_A}
\bar y =\frac{x^2}{c_Dc_a}\simeq \frac{\mu}{a} \left(1-\frac{\sigma^2}{2D}\right)~.
\end{equation}
This is the usual ``mean-field'' singularity close to the (directed percolation) extinction transition
for $\mu\to0$, but with amplitude reduced by $\sigma^2/2D$.

\item{} For $\beta<1$, the linear term must be balanced with the non-analytical correction from $x^{1+2\beta}$,
resulting in $x^{2\beta}\propto c_\mu$. 
Restoring the pre-factors results in a leading singularity of the form
\begin{equation}\label{pos_mean_eq_B}
\bar y =\frac{x^2}{c_Dc_a}\simeq  \frac{\sigma^4}{4aD} \left( \frac{-\Gamma(c_D)}{\Gamma(-c_D)} \frac{\mu}{D} \right)^{\frac{\sigma^2}{2D}}~.
\end{equation}
This is very interesting in that seascape noise with $\sigma^2>2D$ leads to a new a
universality class with the  continuously varying extinction exponent of $(\sigma^2/2D)>1$.

\end{itemize}

\begin{figure}[t]
\centering
\includegraphics[width = 0.5\textwidth]{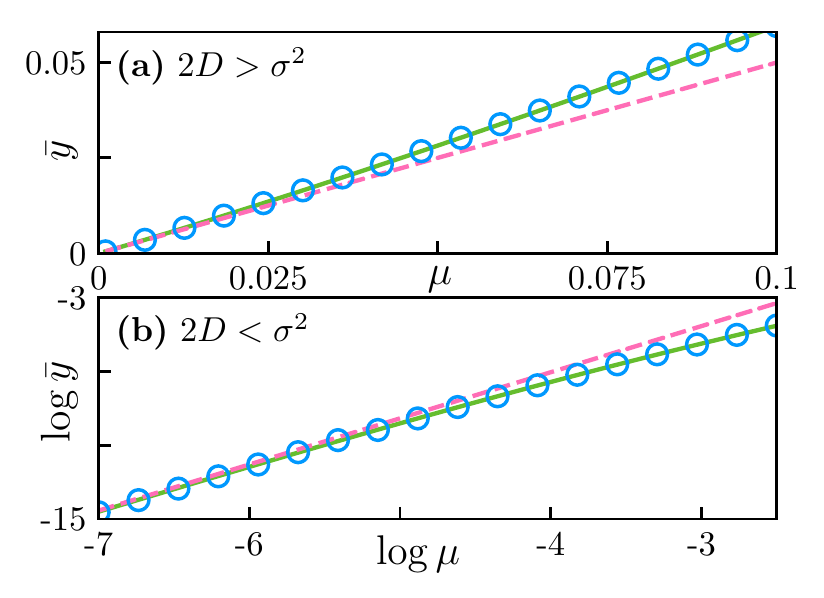}
\caption{Singular behavior close to  extinction as $\mu\to0$.
The solid curve shows the predicted mean via root-finding on Eq.~\eqref{Pos_Mean}, and the circles depict the simulated means obtained from using these chosen values. The dashed curves show the small $\mu$ asymptotics from Eq.~\eqref{pos_mean_small_mu_asym} (exact prefactors given in Appendix~\ref{par_ratio_supsec}). Here, $a=1$, $\sigma = 1$, and $\bar y$ is selected in line with the self-consistency condition of Eq.~\eqref{Pos_Mean}. In top panel (a), $D = 1.0$, so $2D > \sigma^2$, while in (b), $D = 0.2$, so $2D < \sigma^2$. Numerical simulations used $5\times 10^4$ replications with $dt = 0.01$ across 10 time units for (a) and 50 for (b), with the last 2.5 units being used for sampling. }
\label{mean_small_mu_plot}
\end{figure}

\begin{figure}[t]
\centering
\includegraphics[width = 0.5\textwidth]{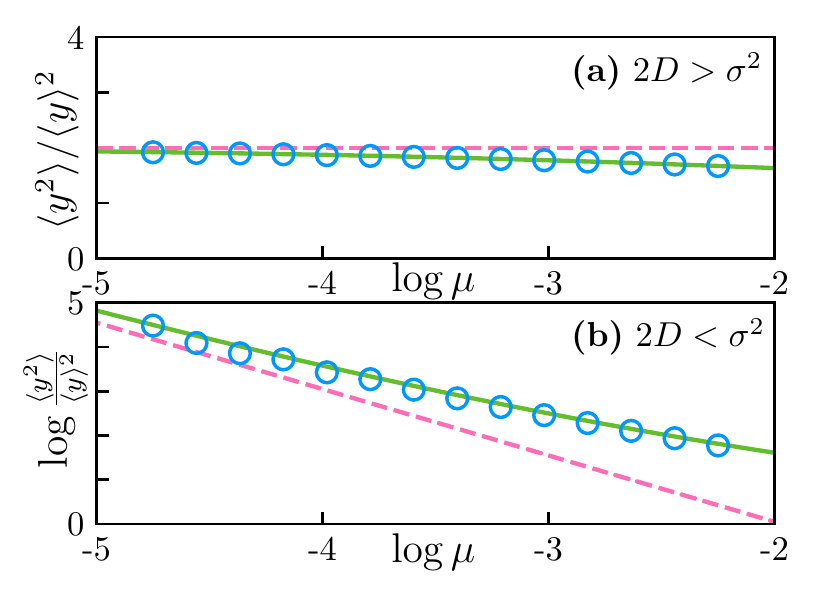}
\caption{The ratio of the second to squared-first moments near the extinction transition $\mu\to0$. The solid curve shows the predicted moment ratio via root-finding on Eq.~\eqref{Pos_Mean}, and the circles depict the simulated ratios obtained from using these chosen values. The dashed curves show the small $\mu$ asymptotics from Eq.~\eqref{pos_ratio_small_mu} (exact prefactors given in Appendix~\ref{par_ratio_supsec}). Here, $a=1$, $\sigma = 1$, and $\bar y$ is selected in line with the self-consistency condition of Eq.~~\eqref{Pos_Mean}. In top panel (a), $D = 1.0$, so $2D > \sigma^2$, while in (b), $D = 0.2$, so $2D < \sigma^2$. Numerical simulations used $2\times 10^6$ replications with $dt = 0.01$ across 10 time units for (a) and 50 for (b), with the last 2.5 units being used for sampling. }
\label{ratio_small_mu_plot}
\end{figure}

\begin{figure}[h]
\centering
\includegraphics[width = 0.5\textwidth]{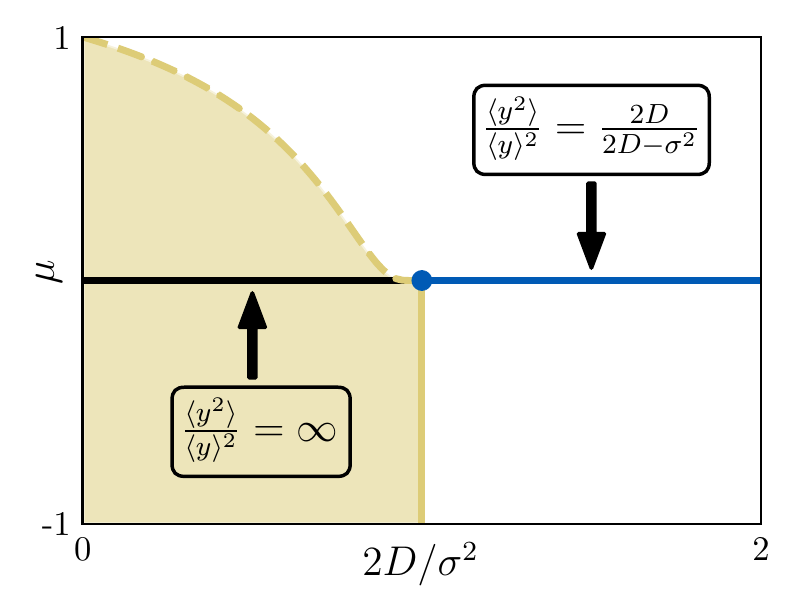}
\caption{Phase diagram of Eq.~\eqref{basic_param} in the large $N$ limit. 
Extinction occurs as $\mu\to0$  (solid blue and black lines), with portions to the right and left of the blue dot exhibiting distinct critical  behaviors.  The vertical line extending from the dot  to $\mu<0$, separates narrow and broad probability distributions. 
Broad distributions extend to the shaded region below the crossover (dashed) line,
here obtained from the condition (see Eq.~\eqref{pos_ratio_small_mu}b)  $\mu^{1 - \sigma^2/(2D)} \geq 1.5$.}
\label{phase_diagram_plot}
\end{figure}

Thus, excluding the exceptional point at $\beta=1$, the singular behavior of the mean is summarized by
\begin{alignat}{1}\label{pos_mean_small_mu_asym}
\bar y \propto \begin{cases}
\mu , & 2D > \sigma^2; \\
\mu^{\sigma^2/(2D)}, & 2D < \sigma^2.
\end{cases}
\end{alignat}
This is confirmed in Fig.~\ref{mean_small_mu_plot}, with one case having linear dependence and the other exhibiting a power-law singularity. 
To gain insight into this unusual scaling, note that averaging Eq.~\eqref{Pos_dyn} over noise, and using $\Exp{y} = \yb$
in the stationary regime, immediately yields 
\begin{equation}\label{eq:yandy2}
 \mu \yb=a \Exp{y^2}=\frac{\int_{0}^{\infty} y^2 \hat \rho(y) dy}{\int_{0}^{\infty}  \hat \rho(y) dy}
 \propto \begin{cases}
\yb^2 , & c_D>1; \\
\yb^{1+c_D}, & c_D<1.
\end{cases}
\end{equation}
As noted before $\hat \rho(y)$ has a power-law form, $y^{-2-c_D}$ (for $\mu\to0$),
which is cutoff by $c_D\yb$ at small $y$ and $(c_a)^{-1}$ at large $y$.
The denominator of Eq.~\eqref{eq:yandy2} is dominated by the short-distance cutoff and scales as $(c_D\yb)^{-1-c_D}$;
while the numerator may be governed by the small- or large-$y$ cutoff depending on whether $1-c_D$ is negative or positive.
In the former case, the numerator scales as $(c_D\yb)^{1-c_D}$, in the latter as $c_a^{c_D-1}$, leading to the
proportionalities indicated in equation~\ref{eq:yandy2}.
Solving for $\yb$ in the two cases then leads to the scalings in Eq.~\eqref{pos_mean_small_mu_asym}.

The ratio $\langle y^2 \rangle/\langle y\rangle^2$ is a measure of the breadth of the distribution.
Keeping track of prefactors, as shown in full in Appendix~\ref{par_ratio_supsec}, yields
\begin{equation}\label{pos_ratio_small_mu} 
\begin{aligned}
\frac{\langle y^2 \rangle}{\langle y \rangle^2} =~& \frac{2D}{2D-\sigma^2},  && 2D > \sigma^2~; \\
\frac{\langle y^2 \rangle}{\langle y \rangle^2} \propto~& \mu^{1 - \sigma^2/(2D)},  && 2D < \sigma^2~.
\end{aligned}
\end{equation}
Note that this behavior  smoothly connects to that obtained previously for $\mu \leq 0$:  For $2D/\sigma^2 > 1$, we obtain the same finite ratio of $\langle y^2 \rangle/\langle y\rangle^2$ on both sides of the transition. For $2D/\sigma^2 < 1$, the moment ratio  diverges as $\mu\to0$, matching the infinite ratio from before. The difference in behavior is clearly confirmed in Fig.~\ref{ratio_small_mu_plot}.  This suggests that parameter space can be divided into two regions: one where the variance is large compared to the mean, and one where they are comparable. The line that separates these regions is set by the ratio of the
coupling $D$ to the seascape noise variance $\sigma^2$, and is schematically shown in Fig.~\ref{phase_diagram_plot}. 
(Distinct singular behaviors for moments or cumulants is typically associated with multifractality, 
and has been observed in critical behavior of systems with quenched disorder~\cite{emig2000thermodynamic, emig2001probability}.)

\section{Demographic noise} \label{dem_sec}

\begin{figure}[h]
\centering
\includegraphics[width = 0.5\textwidth]{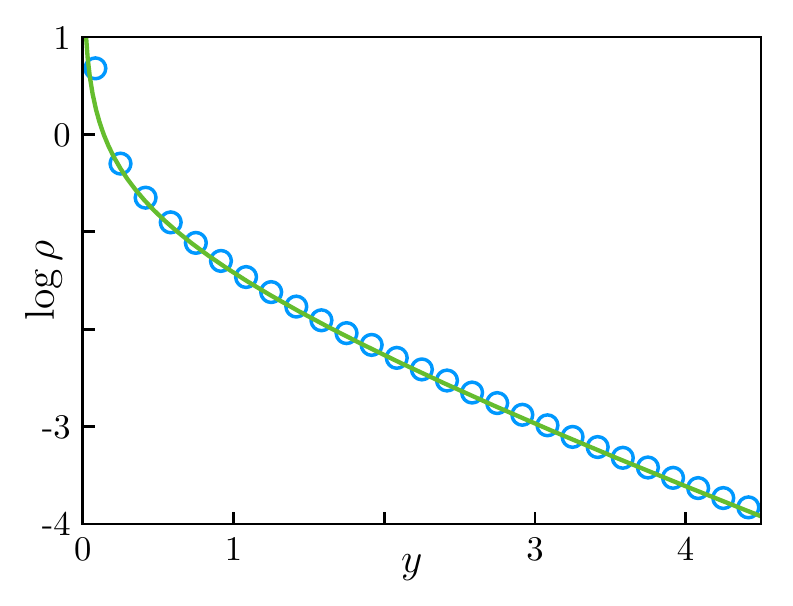}
\caption{Simulated versus predicted probability densities for demographic noise. Circles show simulation results from $5\times 10^6$ runs of Eq.~\eqref{demog_MF_inf_dyn}; the solid line depicts the exact steady--state solution from Eq.~\eqref{demog_MF_pdf}. Simulation was run using stochastic Runge-Kutta, with $y_0 = 1$, $D = 0.25$, $\sigma_2 = 1.0$, $\sigma_d = 0.5$, and $\mu = a = 0$. The time-step of integration was $0.01$, with a total integration time of $25$ units.}
\label{pdf_MF_demog_plot}
\end{figure}

We may well ask if the extinction transition described above  persists with the addition of demographic noise,
i.e. for the equation
\begin{equation}\label{eq:BothNoise}
dy_k =\mu y_k - a{y_k}^2 + D (\bar y - y_k) + \sigma y_k dW_k + \sigma_d \sqrt{y_k} dW'_k~.
\end{equation}
We first consider demographic noise by itself (with $\sigma=0$), and then the mixed case
($\sigma_d\neq0$ and $\sigma\neq0$).

\subsection{Demographic noise with $\mathbf{\mu = 0}$} \label{dem_0_sec}

For $\mu = a = 0$, because of the conserved form of the deterministic part, 
$\bar y=\Exp{y} = y_0$, leading to the stochastic differential equation
\begin{equation}\label{demog_MF_inf_dyn}
dy =D \left(y_0 - y \right) dt + \sigma_d \sqrt{y} ~dW~.
\end{equation}
Then, following the same Fokker-Plank procedure as before, we find the exact distribution for $y$ as 
\begin{equation}\label{demog_MF_pdf}
\rho(y) = \frac{{c'_D}^{c'_D y_0}}{\Gamma(c'_D y_0)} e^{-c'_D y} y^{-1 + c'_D y_0}~,
\end{equation}
with $c'_D = 2D/\sigma_d^2$. This is a so-called Erlang distribution with a shape parameter $k = c'_D y_0$ and rate parameter $\lambda = c'_D$. 

Note that the tail of this distribution at large 
$y$ is exponential, as opposed to the power-law in the case of seascape noise. This leads to well-behaved moments, and the moment ratio
\begin{equation}
\frac{\Exp{y^2}}{\Exp{y}^2} = 1 + \frac{1}{y_0} \frac{\sigma_d^2}{2D}~.
\end{equation}
As an aside, we note that for demographic noise the probability distribution is singular (albeit normalizable) for $y\to 0$ (as may be seen in Fig.~\ref{pdf_MF_demog_plot}). Consequently, negative moments $\Exp{y^{-m}}$ are infinite for $m > y_0 c_D$. 
\subsection{Demographic noise with $\mathbf{\mu < 0}$} \label{dem_neg_sec}

Since the population decays to zero, we can ignore the quadratic term as before, and focus on the
stochastic differential form
\begin{equation}\label{eq:linear}
dy_k = D \bar y(t) - (D+|\mu|)y_k + \sigma_d \sqrt{y_k} dW~,
\end{equation}
with $\bar y(t)=y_0e^{-|\mu| t}$. 
For $D=0$, random fluctuations at a particular location can set $y_k=0$ (an absorbing state). 
At any time $t$, the global population is then composed of a heterogenous set of extinct nodes, 
and nodes with non-zero $y_k$.
The initial probability distribution thus develops a delta-function at the origin that gradually grows
to encompass  the entire probability.
A finite $D$ acts as a source that feeds into the population at each location, removing
the possibility of local extinction (say of an infection). This is despite the fact that this source (from
the global population) is itself decaying exponentially.  Unlike other extinction models (e.g.,~\cite{bittihn2020containment}), 
the overall population decay does not arise from individual sub-populations going extinct, but rather from
all sub-populations decaying at an even pace.  Moreover, while the global decay rate is $|\mu|$, Eq.~\eqref{eq:linear}
 indicates that local nodes relax in response to global and noise fluctuations at a more rapid rate of $(D+|\mu|)$. 

In the limit of $D\gg|\mu|$ we can then take advantage of separation of time scales and
assume that local nodes arrive at a quasi-steady state. 
Ignoring the slow time dependence
of $\bar y(t)$, this quasi-steady distribution is given by Eq.~\eqref{demog_MF_pdf} with
$y_0$ replaced by $\bar y(t)$. 
A typical stochastic node trajectory can be constructed by setting the left hand side of
Eq.~\eqref{eq:linear} to zero; the right hand side admits a positive solution for $\sqrt{y_k(t)}$
for any realization of noise, given $\bar y>0$.

\subsection{Demographic noise with $\mathbf{\mu > 0}$} \label{dem_pos_sec}
For $\mu>0$, we again assume uniformity and use the law of large numbers to assert $\bar y = \Exp y$. This reduces the problem to a single stochastic differential equation
\begin{equation}\label{demog_Pos_dyn}
dy = \{ \mu y - a {y}^2 + D \left(\bar y - y \right) \}dt + \sigma_d \sqrt{y}~dW~.
\end{equation}
The stationary state of the corresponding Fokker-Plank equation satisfies
\begin{equation} \label{demog_Pos_ODE}
0 = y\rho' + (c'_D-c'_\mu)y\rho +(1 - c'_D\bar y)\rho + c'_a y^2 \rho~,
\end{equation}
where  $c'_a = 2a/\sigma_d^2$.  This admits an unnormalized solution of the form 
\begin{equation}\label{demog_Pos_pdf}
\hat \rho(y) = y^{-1+ c'_D \bar y} \exp\left[ -(c'_D - c'_\mu)y - \frac{c'_a}{2} y^2 \right]~.
\end{equation}
While the distribution still diverges as a power-law close to the origin, a finite $\bar y$ renders
the PDF normalizable.

\begin{figure}[t]
\centering
\includegraphics[width = 0.5\textwidth]{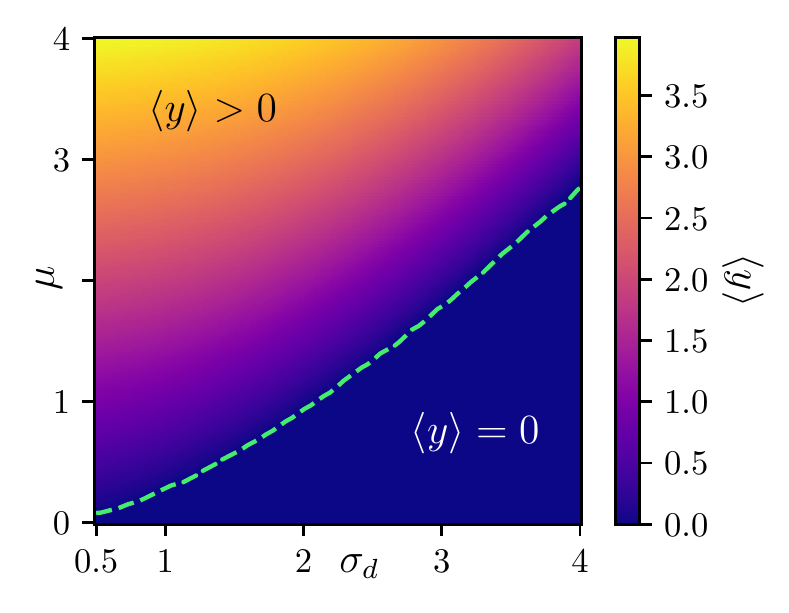}
\caption{`Heat map' depicting the dependence of the self-consistent mean $\yb$ for demographic noise. 
The shade of the color indicates the result for $\bar y$  from Eq.~\eqref{demog_Pos_Mean}. 
The dashed line indicates the critical value of $\mu_c$ below which $\bar y = 0$ exactly. (Rootfinding was done with $a= 1$ and $D = 1.5$).}
\label{demog_heat_plot}
\end{figure}

For computing the normalization and moments of the distribution, we note the identity 
\begin{widetext}
\begin{equation} \label{demog_moments}
\resizebox{0.93\textwidth}{!}{$\int y^m \hat\rho(y) dy = \frac{1}{2}\left(\frac{\sigma_d^2}{a}\right)^{z+\frac{m}{2}} \left[ \Gamma\left(z+\frac{m}{2}\right)\F{z+ \frac{m}{2}, \frac{1}{2}, r^2} + 2 r \Gamma\left(\frac{1+m}{2} + z\right)\F{\frac{1+m}{2} + z, \frac{3}{2}, r^2} \right]$},
\end{equation}
\end{widetext}
where  $r = (c'_\mu - c'_D)/\sqrt{2 c'_a}$ and $z= c'_D \bar y /2$, and ${}_1\mathcal{F}_1$ is the Kummer confluent hypergeometric function. 
While more cumbersome than in the seascape case, the mean can still  be obtained from the self-consistency condition
\begin{equation} \label{demog_Pos_Mean}
\bar y=\frac{\int y \hat\rho(y) dy}{\int  \hat\rho(y) dy}~,
\end{equation}
by numerical and perturbative analysis, which is further detailed in Appendix~\ref{dem_ratio_supsec}.

However, unlike the case of seascape noise, a finite solution for $\bar y$ does not exist for all $\mu>0$.
Instead,  a critical value of $\mu_c$ should be exceeded for a finite mean, as depicted in 
Fig.~\ref{demog_heat_plot}. 
An implicit equation determining $\mu_c$ is provided in the supplementary material, in Appendix~\ref{dem_muc_supsec}.  
In the limit of small noise, it leads to $\mu_c \propto \sigma_d^2 $, while for large noise
$\mu_c  \propto \sigma_d \sqrt{\ln\sigma_d} $.  
To characterize critical behavior near the extinction transition, 
we consider $\mu = \mu_c +\delta\mu$ for $\delta\mu\ll \mu_c$.
As indicated in the supplements, the self-consistency condition of Eq.~\eqref{demog_Pos_Mean}
yields to an analytic expansion in the vicinity of $\mu_c$, which can be rearranged to yield
$\bar y \propto \delta\mu$.
This behavior is confirmed numerically in Fig.~\ref{mean_small_mu_mixed_plot}a.

We may also inquire about the behavior of higher moments. From averaging the equations of 
motion it is easy to check that $\langle y^2\rangle=\mu \langle y\rangle/a$.
It can also be checked that (unlike seascape noise) all moments vanish at criticality in the same manner,
i.e. $\langle y^m\rangle\propto\langle y\rangle\propto (\delta\mu)$.
Consequently, the ratio $\frac{\Exp{y^2}}{\Exp{y}^2}$ diverges as $1/(\delta\mu)$ on approaching $\mu_c$,
as depicted in Fig.~\ref{ratio_small_mu_mixed_plot}a. 

\subsection{Combined demographic and seascape noise} \label{mix_sec}

The novel critical behaviors at extinction observed for seascape noise thus give away to standard behavior
of directed percolation with demographic noise. 
It is thus important to inquire about the nature of the transition when both demographic and seascape noise
are present. Since each noise component has an independent physical motivation,  a mixed--noise model is quite realistic. 
The two noise elements can alternatively be represented by a single noise, which after replacing $\bar y$ for the population average,  
leads to the stochastic differential equation
\begin{alignat}{1}\label{mix_Pos_dyn}
dy =~& \{ \mu y - a {y}^2 + D \left(\bar y - y \right) \}dt \notag \\
&+ \sqrt{{\sigma_d}^2 y + {\sigma}^2 {y}^2} ~dW~.
\end{alignat}

\begin{figure}[t]
\centering
\includegraphics[width = 0.5\textwidth]{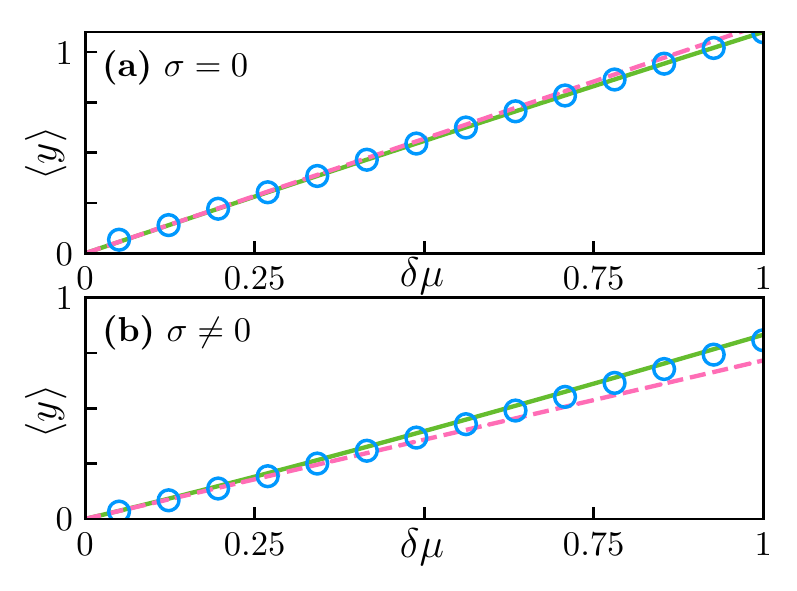}
\caption{Critical behavior close to extinction for both demographic noise 
(Eq.~\eqref{demog_Pos_dyn}) and mixed noise (Eq.~\eqref{mix_Pos_dyn}). 
The solid curve shows the predicted mean via root-finding on the appropriate self-consistency equations, while the circles depict the simulated means from the stochastic differential equations. 
The dashed curves mark the predicted linear dependence close to the transition, with slopes given in appendices~\ref{dem_ratio_supsec} and~\ref{mix_ratio_supsec}. 
Here, $a=1$, $D = 1$, and $\bar y$ is selected to be in line with the appropriate self-consistency condition. In panel (a), $\sigma = 0$ and $\sigma_d = 1$; in (b), $\sigma = 1$ and $\sigma_d = 0.25$. Numerical simulations used $5\times 10^4$ replications with $dt = 0.01$ across 10 time units for (a) and 50 for (b), with the last 2.5 units used for sampling. }
\label{mean_small_mu_mixed_plot}
\end{figure}

\begin{figure}[h]
\centering
\includegraphics[width = 0.5\textwidth]{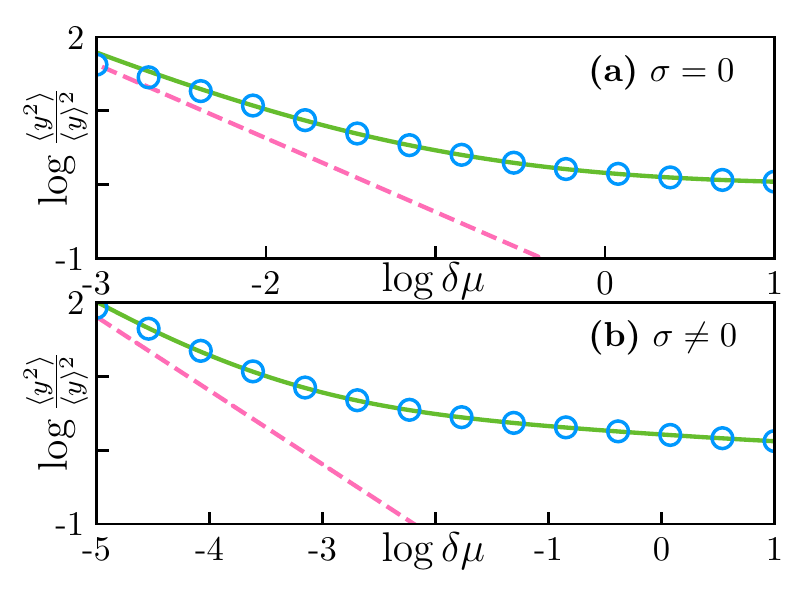}
\caption{Scaling of the moment ratio, $\Exp{y^2}/\Exp{y}^2$, for both demographic noise (Eq.~\eqref{demog_Pos_dyn}) and mixed noise (Eq.~\eqref{mix_Pos_dyn}). The solid curve shows the predicted moments via root-finding on the appropriate self-consistency equations, while the circles depict the simulation results. The dashed curves mark the predicted small $\delta\mu$ scaling, with exponent -1 and prefactors in appendices~\ref{dem_ratio_supsec} and~\ref{mix_ratio_supsec}.  
Here, $a=1$, $D = 1.5$, and $\bar y$ is selected to be in line with the appropriate self-consistency condition. In panel (a), $\sigma = 0$ and $\sigma_d = 1$; in (b), $\sigma = 1$ and $\sigma_d = 0.25$. Numerical simulations used $2\times 10^4$ replications with $dt = 0.01$ across 10 time units for (a) and 50 for (b), with the last 2.5 units being used for sampling. }
\label{ratio_small_mu_mixed_plot}
\end{figure}

Analysis of the corresponding Fokker--Planck equation leads to a stationary distribution proportional to 
\begin{equation}\label{mix_Pos_pdf}
\hat \rho(y) =  y^{-1 + \bar y c_D'} \left(y + \frac{{\sigma_d}^2}{{\sigma}^2} \right)^{-1 - \bar y c_D'+c_\mu-c_D+q}e^{-c_a y}~,
\end{equation}
where $q = 2a{\sigma_d}^2/{\sigma}^4$.
Note that the (integrable) divergence close to zero (proportional to $ y^{-1 + \bar y c_D'}$) is identical that of pure demographic noise.
On the other hand, the exponential decay at large $y$ is the same as that of seascape noise, although the subleading power-law is
modified by an additional power of $y^{1 - \bar y c_D' +q}$. 

Since the extinction transition is dominated by the behavior for $y$ close to zero, its critical behavior turns out to be the same
as that of demographic noise. 
Indeed, $\bar y=\Exp{y}$ vanishes proportionately to $\delta\mu$ (as seen in Fig~\ref{mean_small_mu_mixed_plot}b), 
and the moment ratio, $\Exp{y^2}/\Exp{y}^2$, still scales inversely with $\delta\mu$ (as verified in Fig.~\ref{ratio_small_mu_mixed_plot}b). 
This is shown explicitly in Appendix~\ref{mix_ratio_supsec}.  
Such scaling is regardless of the ratio of $\sigma_d$ to $\sigma$, indicating that any amount of demographic noise will destroy the unusual extinction transition observed in the seascape case. 
Thus the novel universality and broad distributions, characteristic of seascape noise close to an extinction transition, disappear
in the presence of (expected) demographic noise.

\section{Discussion} \label{discussion_sec}

Both seascape fluctuations and demographic noise have well defined origins. The former arises from natural variations in population fitness and access to resources, whereas the latter is the result of moving from a discrete to a continuum description. Both are unavoidable in real world systems, and neglecting one or the other can cause drastically incomplete predictions of the breadth of a population. 
In particular, seascape fluctuations broaden the tail of the distribution at large values,
while demographic stochasticity controls its behavior close to zero.

To understand  the importance of seascape fluctuations, it is instructive to
consider Eq.~\eqref{basic_param_space} in the limit of $a=0$ and $D=0$. 
At each point $\ln y$ now performs a random walk, resulting in broad log-normal distributions
for which characterizing the population in terms of mean $\Exp{y}$ and variance $\Exp{y^2}_c$
is inadequate. 
A finite diffusion coefficient (still with $a=0$) is expected to modify the log-normal distribution.
The resulting behavior, in the universality class of directed polymers in random media (DPRM), is 
 highly dependent on dimensionality: In one dimension it leads to the celebrated Tracy--Widom
 distribution~\cite{dotsenko2010bethe, quastel2015one, chu2016probability}, while in dimensions larger than 2, a critical value of $D_c$ separates
 broad and narrow distributions~\cite{halpin20122+, kardar1987scaling, halpin2013extremal}.
 To circumvent difficulties associated with spatial dimensionality, in this paper we consider a
 mean-field limit (Eq.~\eqref{basic_param}) in which the parameter $D$  accounts for
 migration between any pair of sites, irrespective of their separation. 
In this case, seascape noise of strength $\sigma$ generates a power-law tail in the distribution for
large populations which falls off with exponent $2+2D/\sigma^2$. 
Much like the case of DPRM at high dimensions, this signals a transition from broad to a narrow 
 distributions for $D>D_c=\sigma^2/2$. 
 
 For $a=0$, a finite $\mu$ simply makes the mean, $\yb$, time dependent (proportional to $e^{\mu t}$)
 without modifying the shape of the distribution.  For $\mu>0$, a finite $a$ is needed to curtail
 the exponential growth of the population; a self-consistency requirement now enables
 computing a stationary $\bar{y}$. 
 The finite $a$ also leads to an exponential decay of the distribution at large values, resulting in well-behaved moments.
The most interesting aspect of the problem is then the nature of the extinction transition as $\mu\to0$:
For $D>D_c$, the mean (and all other moments) vanish linearly with $\mu$, as expected for
a mean-field extinction transition. 
However, for $D<D_c$ the mean vanishes as $\mu^{\sigma^2/(2D)}$, with higher moments
vanishing with other exponents (harking back to the broad distribution expected for $\mu=0$).
 
 While the above unconventional critical behavior is an accurate description of Eq.~\eqref{basic_param}
 as $\mu\to0$, its relevance to an actual species extinction must also account for demographic stochasticity.
 For $D=0$, demographic noise in Eq.~\eqref{basic_demog} is incompatible with a continuous probability distribution, with fluctuations causing local extinctions resulting in a growing delta-function in the PDF
 at the absorbing point $y=0$. 
 A finite $D$ removes the delta-function by seeding new population at all sites through migration from other sites, resulting in the steady state distribution of Eq.~\eqref{demog_MF_pdf}, with a normalizable
 divergence at the origin. 
 (For $|\mu|\ll D$, the same distribution, albeit with a time dependent $\bar{y}(t)$ can be
 used to describe growing or shrinking populations.)
 However, the boundary between growth and extinction is no longer at $\mu=0$ in the presence
 of demographic noise, and the self-consistency condition should be used to identify a critical
 value of $\mu_c$. 
 The nature of the extinction transition at $\mu_c$ is now conventional, belonging to the expected
 directed percolation universality class. 
Finally, we find that when both seascape and demographic noise are present, the resulting extinction
transition (at a finite $\mu_c$) is again conventional. This is not too surprising as extinction 
characterizes behavior when $y\to 0$, where demographic fluctuations are paramount.

The overarching theme of this investigation is the emphasis on the need to characterize 
the full distribution of a population and not just its mean.
In the course of this investigation, in the mean-field limit, we have encountered  a few distributions of some repute: For seascape noise, a scaled inverse chi-squared distribution appears; for demographic noise an Erlang shows up, and when both noises are mixed (for $\mu = a = 0$), a rescaled beta prime distribution manifests. 
The special functions characterizing these distributions are typically solutions of relatively simple
differential equations. In our context, these equations appear naturally in computing steady states of
Fokker-Planck equations, providing examples of their applicability.

The covid-19 epidemic has motivated myriad current studies of spread and control of epidemics~\cite{giordano2020modelling, tuite2020mathematical, chatterjee2020healthcare, kucharski2020early}.
With notable exceptions (e.g. Ref.~\cite{bittihn2020containment}), most studies follow a number of 
time-dependent quantities that are implicitly assumed to characterize the whole distribution. 
The approach of this paper can be used to examine actual distributions via the self-consistency
condition in the mean-field limit, for multiple stochastic equations, generalizing Eq.~\eqref{mix_Pos_dyn}. 
Even for a single quantity, we may  generalize the exponent of the stabilizing nonlinear term from 2 to 
an arbitrary $\alpha$. This is not without motivation, since instead of an embedded logistic equation at every point in space, we would now have a {\it Richards' equation}. The Richards' equation acts as a generalization of logistic growth, and has been used in a variety of ecological models~\cite{richards1959flexible, fekedulegn1999parameter}. 
Indeed, most results presented in this paper have natural generalizations for $\alpha >1$, 
including corresponding distributions. 
The most notable complexity arises in the self-consistency equation for the mean,
which requires numerical analysis and lacks closed forms. 

Another interesting query is the applicability of the mean-field results to the finite dimensional
extinction problem. There is a certain richness of dimensional dependence in directed percolation models, 
but we may nonetheless expect some qualitative behavior in the mean-field case to carry over to finite dimensions~\cite{korolev2010genetic}. For example, in the important cases of one and two dimensions, the limiting distribution for $a=0$ 
arising from Eq.~\eqref{basic_param_space} is always broad. It would be interesting
to investigate the nature of the extinction transition with seascape noise (with and without
demographic noise) in these dimensions.

Finally, we note that certain predictions in this model may be experimentally replicable. Well-mixed microbial populations are often represented via a mean-field migration model, which we employ here. 
While tracking entire distributions is cumbersome (albeit not impossible), the mean and standard deviation of the population should be relatively easy to track. 
Our model predicts that the ratio of these quantities makes a dramatic change, passing from finite
to very large as the strength of seascape noise is increased above a threshold value.
The fitness of a population can be controlled via the resources afforded to the microbes, 
and local fluctuations in this quantity can be constructed by randomizing access to resources
(e.g. light levels). 
While the results presented in this paper may not be robust to various perturbations, 
an experiment probing the effects of seascape fluctuations in fitness should yield highly
informative results.

This research was supported by a McDonnel fellowship to Bertrand Ottino-L\"{o}ffler, as well as by NSF through grants \#~DMR-1708280 and \# PHY-2026995 (MK).

\bibliography{SeaScape_Bib}{}

\clearpage

\appendix

\section{Moments for seascape noise} \label{par_ratio_supsec}

We will elaborate upon the case of seascape noise with positive fitness, that is,
\begin{equation} \label{par_dyn_sup}
dy = \left[ \mu y - ay^2 + D \left(\bar y - y \right) \right]dt + \sigma y dW\,,
\end{equation}
with $\mu > 0$. In the main body of the paper we find that, in the appropriate limit, the distribution of $y$ is proportional to 
\begin{equation}\label{par_pdf_sup}
\hat \rho(y) = e^{-c_D \yb /y} e^{-c_a y} y^{-2 - \beta},
\end{equation}
where we used the law of large numbers to assert $\bar y = \langle y \rangle$. We further define $c_D = 2D/\sigma^2$, $c_a = 2a/\sigma^2$, and $\beta = 2(D - \mu)/\sigma^2$. 

In order to get the right normalization $Z$ for the distribution, and to choose a self-consistent value of $\yb$, we need to use the unnormalized moments 
\begin{equation}\label{par_moments_sup}
\int_{0}^\infty y^m \hat \rho(y) dy = 2 \left( \frac{c_a}{c_D \yb} \right)^{(1+\beta)/2} \mathcal{K}_{\beta+1}(2 \sqrt{c_D c_a \yb}).
\end{equation} 
Therefore, we have that 
\begin{equation} \label{par_mean_sup}
\sqrt{\frac{c_a}{c_D} \yb} = \frac{\K_\beta(2 \sqrt{c_D c_a \yb})}{\K_{\beta+1}(2 \sqrt{c_D c_a \yb})}\,,
\end{equation}
as the self-consistent restriction on  $\Exp{y} = \yb$.

Because we have assumed that $\yb = \Exp{y}$ and $\yb$ is constant in time, then we can take the expectation of both sides of the dynamics to get $\mu \Exp{y} =  a \Exp{y^2}$.  Therefore, we have that 
\begin{equation}\label{generic_ratio_sup}
\frac{\Exp{y^2}}{\Exp{y}^2} = \frac{\mu/a}{\Exp{y}}.
\end{equation}
So if we want to understand the ratio $\Exp{y^2}/\Exp{y}^2$ in the extinction limit, that is, as $\mu \to 0$ and $\yb \to 0$, then all we need to do is understand  the asymptotics of~\eqref{par_mean_sup} alone.

For the sake of brevity we will define $x = \sqrt{c_D c_a \yb}$. The most convenient form of $\K$, the modified Bessel function of the second kind, is
\begin{alignat}{2}
\K_\beta(2x) =&&~& \frac{\pi/2}{\sin(\pi\beta)} \sum_{m=0}^\infty \frac{x^{2m-\beta}}{m!\Gamma(m-\beta+1)}  \notag \\
&& - &  \frac{\pi/2}{\sin(\pi\beta)} \sum_{m=0}^\infty  \frac{x^{2m+\beta}}{m! \Gamma(m+\beta+1)}. \label{bess_def}
\end{alignat}
Thus, limiting to the lowest order terms in $x$, the right hand side of Eq.~\eqref{par_mean_sup} becomes
\begin{widetext}
\begin{alignat}{1}
\frac{\K_\beta(2x)}{\K_{\beta+1}(2 x)} \simeq  (-1) &  \left( \frac{x^{-\beta}}{\Gamma(-\beta+1)} + \frac{x^{-\beta+2}}{\Gamma(-\beta+2)} - \frac{x^{\beta}}{\Gamma(\beta+1)} - \frac{x^{\beta+2}}{\Gamma(-\beta+2)} \right) \notag \\ 
& \cdot \left( \frac{x^{-\beta-1}}{\Gamma(-\beta)} + \frac{x^{-\beta+1}}{\Gamma(-\beta+1)} - \frac{x^{\beta+1}}{\Gamma(\beta+2)} - \frac{x^{\beta+3}}{\Gamma(\beta+3)} \right)^{-1}.  \label{par_mean_ap_sup}
\end{alignat}
\end{widetext}

Disambiguating which terms are actually most relevant depends on the choice of $\beta$, with there being three regimes of note.  We shall elaborate on each of these in turn.   

\subsubsection{$ \beta > 1$}
In this case, the $x^{-\beta}$ and $x^{-\beta+2}$ terms dominate the numerator of ~\eqref{par_mean_ap_sup}, whereas the denominator is dominated by the 
$x^{-\beta-1}$ and $x^{-\beta+1}$ terms. Therefore, we find that  
\begin{equation} \label{bess_ratio1_sup}
\frac{\K_\beta(2x)}{\K_{\beta+1}(2 x)} \simeq \frac{x}{\beta} \left(1 + \frac{x^2}{\beta(1-\beta)} \right). 
\end{equation}

Since $x \propto \sqrt{\yb}$, the $\sqrt{\yb}$ on the left hand side of Eq.~\eqref{par_mean_sup} cancels out the leading $x$, hence the need to include the second order correction. From here, we can easily solve for $\yb$ to get 
\begin{equation}\label{par_mean_a1_sup}
\yb \simeq \frac{\mu}{D} \frac{\beta(\beta-1)}{c_{D}^2}.
\end{equation}

Expanding $\beta$ in the small-$\mu$ limit then leads to
\begin{equation}\label{par_mean_b1_sup}
\yb \simeq \frac{\mu}{a}\left(1 - \frac{\sigma^2}{2D} \right), 
\end{equation}
which is  linear in $\mu$, as  stated in the main text.  Moreover, we get  
\begin{equation}\label{par_ratio_b1_sup}
\frac{\Exp{y^2}}{\Exp{y}^2} = \frac{2D}{2D - \sigma^2} + O(\mu), 
\end{equation}
which is consistent with the results from the $\mu = 0$ analysis for $2D > \sigma^2$.

\subsubsection{$-1 < \beta < 1$}

Here, the numerator of~\eqref{par_mean_ap_sup} is dominated by the $x^{\beta}$ and $x^{-\beta}$ terms, and the only term from the denominator that survives is the $x^{-\beta-1}$ term. So 
\begin{equation} \label{bess_ratio2_sup}
\frac{\K_\beta(2x)}{\K_{\beta+1}(2 x)} \simeq \frac{x}{\beta} \left(1 + \frac{\Gamma(-\beta)}{\Gamma(\beta)} x^{2\beta} \right). 
\end{equation}
The leading $x$ cancels out the $\yb$ on the left hand side of~\eqref{par_mean_sup}, leving us with the mean being set by 
\begin{equation} \label{par_mean_a2_sup}
\yb \simeq \frac{1}{c_ac_D} \left( \frac{-\Gamma(\beta)}{\Gamma(-\beta)} \frac{\mu}{D} \right)^{1/\beta}. 
\end{equation}
In the small $\mu$ limit, $\beta \to c_D$, so the lowest order behavior in $\mu$ is given by
\begin{equation} \label{par_mean_b2_sup}
\yb \simeq \frac{1}{c_ac_D} \left( \frac{-\Gamma(c_D)}{\Gamma(-c_D)} \frac{1}{D} \right)^{1/c_D} \mu^{1/c_D}. 
\end{equation}
So if $|\beta| < 1$, the mean goes as $\mu^{\sigma^2/(2D)}$.  Moreover, the moment ratio is  given by 
\begin{equation} \label{par_ratio_b2_sup}
\frac{\Exp{y^2}}{\Exp{y}^2} \simeq {c_D}^{1+1/c_D} \left(\frac{-\Gamma(-c_D)}{\Gamma(c_D)} \right)^{1/c_D} {c_\mu}^{1 - 1/c_D}. 
\end{equation}
This provides all the prefactors negelected in the main body Eq.~\eqref{pos_ratio_small_mu}.

\subsubsection{$\beta < -1$}
For this region, the numerator of ~\eqref{par_mean_ap_sup} is dominated by the $x^\beta$ and $x^{\beta+2}$ terms, and the denominator is dominated by the $x^{\beta+1}$ and $x^{\beta+3}$ terms. So
\begin{equation}\label{bess_ratio3_sup}
\frac{\K_\beta(2x)}{\K_{\beta+1}(2 x)}  \simeq \frac{-(\beta+1)}{x} \left(1 + \frac{x^2}{(\beta+1)(\beta+2)} \right). 
\end{equation}
As before, the second order expansion is required.  Solving for $\yb$ gives us 
\begin{equation} \label{par_mean_a3_sup}
\yb = \frac{-(\beta+1)(\beta+2)}{c_a(\beta+2 + c_D)},
\end{equation}
and therefore
\begin{equation} \label{par_ratio_a3_sup}
\frac{\Exp{y^2}}{\Exp{y}^2} = \frac{\beta}{\beta+1} \left(1 + \frac{c_D}{\beta+2} \right)^2.
\end{equation}

Unlike the previous two cases, there is no need to do a small-$\mu$ analysis here. The only way for $\beta = c_D - c_\mu$ to be negative is for $\mu$ to be greater than $D$.  This is not possible in the limit of small $\mu$, so this regime is irrelevant for our purposes.  

\vspace{0.40cm}

\section{Moments for demographic noise} \label{dem_ratio_supsec}

Here, we will show the extinction behavior of our mean-field model under demographic noise. In particular, we will look at 
\begin{equation}\label{Model_demog_sup}
dy_k = \left[ \mu y_k - a {y_k}^2 + D \left(\bar y - y_k \right) \right]dt + \sigma_d \sqrt{y_k} dW_k.
\end{equation}
As before, we use the law of large numbers to assert that the ensemble average, spatial average, and initial condition are identical. If this holds true, then there is some stationary density $\rho$ that is identical for every $y_k$, and we can transform the above equation into a Fokker-Plank equation, where
\begin{equation} \label{dem_PDE_sup}
0 = - \partial_t \rho = \partial_y \left\{ \left[ \mu y - a y^2 + D(\bar y - y) \right] \rho \right\}.
\end{equation}
This simplifies to 
\begin{equation} \label{dem_ODE_sup}
0 = y \rho' + (1 - c_D \yb) \rho + (c_D - c_\mu) y \rho + c_a y^2 \rho,
\end{equation}
with $c_D' = 2D/{\sigma_d}^2$, $c_\mu' = 2\mu/{\sigma_d}^2$, and $c_a' = 2a/{\sigma_d}^2$ as usual. This admits an unnormalized solution of the form 
\begin{equation} \label{dem_pdf_sup}
\hat \rho(y) = y^{-1+c_D' \yb} \exp\left[ -(c_D' - c_\mu') y - (c_a'/2) y^2 \right].
\end{equation}

To get the moments of $y$, it is expedient to define $z = D \yb/{\sigma_d}^2$ and $r = (-D + \mu)/\sqrt{a {\sigma_d}^2}$. Using these definitions, we find that 
\begin{widetext}
\begin{alignat}{1}
\yb &= \frac{\int y \hat \rho(y) dy}{\int \hat \rho(y) dy} = \sqrt{\frac{2}{c_a'}} \frac{\Gamma\left(\frac{1}{2} + z\right) \F{\frac{1}{2}+z, \frac{1}{2}, r^2} + 2 r \Gamma(1+z) \F{1+z, \frac{3}{2}, r^2} } {\Gamma(z)\F{z, \frac{1}{2}, r^2} + 2r \Gamma\left(\frac{1}{2} + z\right) \F{\frac{1}{2}+z, \frac{3}{2}, r^2}},  \label{mean_dem_eq_sup}
\end{alignat}
\end{widetext}
where $\Fz$ is the Kummer confluent hypergeometric function.

We want our analysis to take place near $\mu = \mu_c$, so $\yb$ should be close to zero.  The most practical way to analyze the extinction behavior is to look at both the numerator and denominator of the expression~\eqref{mean_dem_eq_sup} for $\yb$ in the limit of small $z$, since if $\yb$ is small, then so is $z \propto \yb$. Specficially, we have 
\begin{equation} \label{dem_mean1_sup}
\yb \simeq \sqrt{\frac{2}{c_a'}} \frac{A + Bz}{E z^{-1} + F}\,,
\end{equation}
where $A$, $B$, $E,$ and $F$ are constants obtained by expanding in $z$, with $A$ and $B$ given by 
\begin{alignat}{1}
A =~& e^{r^2} \sqrt{\pi}(1+\mbox{Erf}(r)),  \label{dem_A} \\ 
B =~& -e^{r^2} \gamma_E \sqrt{\pi} \mbox{Erf}(r) + \sqrt{\pi}\Fz^{(1,0,0)}\left(\frac{1}{2}, \frac{1}{2}, r^2\right) \notag \\
&+ \sqrt{\pi} \Gamma_P(0, 1/2) e^{r^2} + 2 r \Fz^{(1, 0, 0)}\left(1, \frac{3}{2}, r^2\right), \label{dem_B} 
\end{alignat}
and $E$ and $F$ given by 
\begin{alignat}{1}
E =~& 1, \label{dem_E} \\
F =~&  -\gamma_E + \pi \mbox{Erfi}(r) + \Fz^{(1,0,0)}\left(0, \frac{1}{2}, r^2\right), \label{dem_F}
\end{alignat}
with $\Gamma_P$ being the polygamma, Erf the error function, Erfi the imaginary error function, and $\gamma_E \approx 0.577$ is the Euler-Mascheroni constant. 

Using these approximate forms, it is easy to see that 
\begin{equation} \label{dem_mean2_sup}
\yb = \frac{-2}{c_D'} \frac{c_D' A - \sqrt{2 c_a'} E}{c_D' B - \sqrt{2 c_a'}F}.
\end{equation}
Let's assume that $\mu_c$ is the critical value of $\mu$ where $\yb$ goes from zero to nonzero. 
The above form allows us to obtain the critical $\mu$, since it implies that $\mu_c$ must satisfy  $c_D' A = \sqrt{2 c_a'}$, or 
\begin{equation} \label{dem_muc_sup}
\sqrt{\frac{a}{\pi}} \frac{\sigma_d}{D} = e^{r_c^2} \left( 1 + \mbox{Erf}(r_c)\right)\,,
\end{equation}
with $r_c = (-D + \mu_c)/\sqrt{a {\sigma_d}^2}$.

If we are instead interested in the behavior of the mean near zero, then we can substitute $\mu = \mu_c+ \delta\mu$ for infinitesimal $\delta\mu$, and find
\begin{equation} \label{dem_mean3_sup}
\yb = \frac{-2 \delta A}{c_D' B - \sqrt{2 c_a'} F} \delta \mu,
\end{equation}
where
\begin{equation} \label{dem_dA_sup}
\delta A = \frac{2}{\sqrt{a}\sigma_d} - \frac{2\pi (D - \mu_c)}{a {\sigma_d}^2} e^{{r_c}^2} \left[ 1 - \mbox{Erf} \left( -r_c \right) \right].
\end{equation}

This means that our predicted mean for demographic noise scales as $\yb \propto \delta\mu = \mu - \mu_c$ near extinction, with a slope that is exactly predictable from the above equation.  

By taking the expectation of both sides of the dynamics, we attain $\Exp{y^2} = (\mu/a) \Exp{y}$.  So if we plug this into $\Exp{y^2}/\Exp{y}^2$ and only keep the lowest order terms in $\delta \mu$, we get 
\begin{equation} \label{dem_ratio_sup}
\frac{\Exp{y^2}}{\Exp{y}^2} = \frac{\mu_c + \delta \mu}{a \yb} \propto \frac{1}{\delta \mu}.
\end{equation}
Therefore, unlike in the case of seascape noise, the moment ratio always diverges as $1/\delta \mu$.

\section{$\mu_c$  for demographic noise} \label{dem_muc_supsec}

Here we will find approximate forms for the critical fitness $\mu_c$, at which the mean population goes from zero to nonzero.  
In the prior Appendix, we demonstrated that Eq.~\eqref{dem_muc_sup}  sets the condition for $\mu_c$.  If we let $r_\mu = \mu_c/(\sigma_d\sqrt{a})$ and $r_D = D/(\sigma_d \sqrt{a})$, then this condition can be written as 
\begin{equation}\label{dem_muc_alt_sup}
\frac{1}{r_D \sqrt{\pi}} = e^{(r_\mu - r_D)^2} (1 + \mbox{Erf}(r_\mu - r_D)),  
\end{equation}
where Erf is the error function.  
In the following we examine separately the large and small noise limits of the above equation.

\subsubsection{$\sigma_d \gg 1$}

Since $r_D \propto 1/\sigma_d$, we might expect this quantity to be small.  Let us assume that $r_\mu$ is large, and futher assume that $r_\mu r_D $ is small.  These conditions can be checked for self-consistency at the end. 

Taking these assumptions in hand, we have 
\begin{alignat}{1}
\frac{1}{\sqrt{\pi}r_D} =~& e^{(r_\mu - r_D)^2} (1 + \mbox{Erf}(r_\mu - r_D) ) \notag \\
 \simeq & 2e^{r_{\mu}^2}. \label{dem_muc_alt1_sup}
\end{alignat}
By assumption, we were allowed to neglect the $e^{-2 r_\mu r_D}$ term, as well as approximate the error function by a constant. We now obtain $r_\mu \simeq \sqrt{-\ln (2 \sqrt{\pi} r_D)}$, and therefore get 
\begin{equation} \label{dem_muc1_sup}
\mu_c \simeq \sigma_d \sqrt{- \ln\left(2 \sqrt{\frac{\pi}{a}} \frac{D}{\sigma_d} \right) }.
\end{equation}
Notice that $r_\mu \propto \sqrt{\ln(\sigma_d)}$ and $r_\mu r_D \propto \ln(\sigma_d)/\sigma_d$, making our initial assumptions self-consistent.  

\subsubsection{$0 < \sigma_d \ll 1$}

An efficient way to rewrite Eq.~\eqref{dem_muc_alt_sup} is 
\begin{equation} \label{dem_muc_alt2_sup}
\frac{1}{r_D\sqrt{\pi}} = \mbox{Erfcx}(r_D - r_\mu),
\end{equation}
in terms of the scaled complementary error function, $\mbox{Erfcx}$.  
Since $\sigma_d$ is small, therefore $r_D$ is necessarily large.  If we assume $r_\mu$ is small, then we can take the first order Taylor series of  Erfcx to get 
\begin{alignat}{1}
r_\mu =~& \frac{ -(\sqrt{\pi} r_D)^{-1} + \mbox{Erfcx}(r_D)}{\mbox{Erfcx}'(r_D)} \notag \\
 =~& \frac{ -(\sqrt{\pi} r_D)^{-1} + \mbox{Erfcx}(r_D)}{-2/\sqrt{\pi} + 2r_D \mbox{Erfcx}(r_D)}  \notag \\
 =~& \frac{1}{2r_D}. \label{dem_muc_alt3_sup}
\end{alignat}
Since $r_D$ is large, therefore $r_\mu$ is small and self-consistent, leading to  
\begin{equation}\label{dem_muc2_sup}
 \mu_c = a \frac{\sigma_{d}^2}{2D}.
 \end{equation}
 
Another way to prove this case is by using the fact that $c_D'$ is large, so we can approximate the unnormalized
 distribution~\eqref{dem_pdf_sup} by
\begin{equation}\label{dem_pdf_approx_sup}
\hat\rho(y) \simeq  y^{-1+c_D' \yb} e^{-c_D' y}.
\end{equation}
The second moment is exactly solvable here, with $\Exp{y^2} = \yb^2 + \yb/c_D'$.  If we then take the expectation of both
 sides of the dynamics~\eqref{Model_demog_sup} and plug in the second moment we find that 
\begin{equation} \label{demog_mean_ode_sup}
 \partial_t \yb  = \left(\mu - \frac{a}{c_D'}\right) \yb - a \yb^2.
\end{equation} 
This suggests that $\yb$ converges to zero whenever $\mu < a/c_D'$, which retrieves the 
equation for $\mu_c$~\eqref{dem_muc2_sup}.

So for small noise, the critical fitness scales as $\sigma_{d}^2$, but for large noise it scales as $\sigma_d\sqrt{\ln(\sigma_d)}$, as  described in the main body of the paper.

\section{Moments for mixed noise} \label{mix_ratio_supsec}

Here, we look at the case with mixed seascape and demographic noise, that is, 
\begin{equation}\label{Model_mixed_sup}
dy = \left[ \mu y - a {y}^2 + D \left(\bar y - y \right) \right]dt + \sqrt{ \sigma^2 y^2 + \sigma_{d}^2 y } dW,
\end{equation}
where $\sigma$ and $\sigma_d$ are the respective amplitudes of  seascape and demographic noises. As always, we assume uniform initial conditions and the existence of a stationary distribution $\rho$ with a self-consistent mean $\Exp y = \yb$. By transforming the above into a Fokker-Plank equation, we find the unnormalized stationary distribution 
\begin{equation} \label{mix_pdf_sup}
\hat \rho(y) = e^{-c_a y} y^{-1 + z'} (y + s)^{-1-t-z'},
\end{equation}
where $c_a = 2a/\sigma^2$, $s = \sigma_{D}^2/\sigma^2$,   $z' = 2D\yb/{\sigma_d}^2$, $q =  2a\sigma_{d}^2/\sigma^4$, and $t  = 2(D - \mu)/\sigma^2 - q$. 

In order to get the self-consistent mean, we need to use the following equation
\begin{widetext}
\begin{equation} \label{mix_mean1_sup}
\yb = \frac{s \sin(\pi t)}{\pi} \frac{\Gamma(t) \Gamma(1+z') \F{1+z', 1-t, q} + q^t \Gamma(-t)\Gamma(1+t+z')\F{1+t+z', 1+t, q}} {-\Gamma(z') \FReg{z', -t, q} + q^{1+t} \Gamma(1+t+z') \FReg{1+t+z', 2+t, q} },
\end{equation}
\end{widetext}
with $\Fz$ being the Kummer hypergeometric function and $\FRegz$ being its regularized version. In the extinction limit, $\yb$ must be small, and so $z'$ should be too. In such a limit, we find 
\begin{equation} \label{mix_mean2_sup}
\yb \simeq \frac{s \sin(\pi t)}{\pi} \frac{A' + B'z'}{E'/z' + F'},
\end{equation}
which simplifies into
\begin{equation} \label{mix_mean3_sup}
\yb \simeq \frac{-\sigma_{d}^2}{2D} \frac{2D \sin(\pi t) A' - \pi \sigma^2 E'}{2D \sin(\pi t) B' - \pi \sigma^2 F'}\,,
\end{equation}
where 
\begin{alignat}{1}
A' =~&  -te^{q}q^t  \Gamma(-t)\Gamma(t) + e^{q}q^t  \Gamma(-t)\Gamma(1+t) \notag \\
&+ t e^{q}q^t \Gamma(t)\Gamma(-t, q),  \label{mix_A_sup} \\
B' =~& t e^{q} \gamma_E \Gamma(t) q^t \Gamma(-t) - t e^{q}\gamma q^t \Gamma(t)\Gamma(-t, q) \notag \\
&+ \Gamma(t){\FRegz}^{(1,0,0)}(1, 1-t, q) \notag \\
&+ e^{q}q^t \Gamma_P(0, 1+t) \Gamma(-t)\Gamma(1+t) \notag \\
&+ q^t\Gamma(-t)\Gamma(1+t){\FRegz}^{(1,0,0)}(1+t,1+t,q), \label{mix_B_sup} \\
E' =~& -1/\Gamma(-t), \label{mix_E_sup} \\
F' =~& \gamma/\Gamma(-t) + \int_{0}^{q} u^t e^u du - {\FRegz}^{(1,0,0)}(0, -t, q), \label{mix_F_sup}
\end{alignat}
where $\gamma_E \approx 0.577$ is the Euler-Mascheroni constant, $\Gamma_P$ is the polygamma, and $\Gamma(x, y)$ is the incomplete gamma function $\int_{y}^\infty u^{x-1} e^{-u} du$. 

We can then assert that there is some $\mu_c$ (with corresponding $t_c$) where the mean transitions from zero to nonzero. Given this, and assuming $\mu = \mu_c + \delta\mu$, with $c_{\delta \mu} = 2\delta\mu/\sigma^2$, we find for small $\delta\mu$ that
\begin{equation} \label{mix_mean4_sup}
\yb \simeq  \frac{  \pi  \delta E' + \pi c_D \cos(t_c \pi) A' -  c_D \sin(\pi t_c)\delta A'}{c_{D}'  \sin(\pi t) B' - \pi c_{D}' F'} c_{\delta\mu},
\end{equation}
where $c_D = 2D/\sigma^2$, $c_D' = 2D/\sigma_{d}^2$,  
\begin{alignat}{1}
\delta A' =~& -e^{q} q^{t_c} \left\{ \Gamma(t_c)\Gamma(-t_c, q) \vphantom{\Gamma^{(1,0)}} \right. \notag \\
& +  \Gamma(t_c)\Gamma(-t_c, q)\left[t_c\ln(q) + t_c\Gamma_P(0,t_c)\right] \notag \\
&-\Gamma(-t_c)\Gamma(t_c) \left[1 + t_c\Gamma_P(0,t_c \right] \notag \\
& +  \Gamma(-t_c)\Gamma(t_c) t_c\Gamma_P(0,1+t_c) \notag \\
&\left. -t_c\Gamma(t_c)\Gamma^{(1,0)}(-t_c, q)\right\}, \label{mix_dA_sup} 
\end{alignat}
and 
\begin{equation} \label{mix_dE_sup}
\delta E' = \Gamma_P(0, -t_c)/\Gamma(-t_c).
\end{equation}
In particular, this means that near extinction, the mean scales as $\yb \propto \mu - \mu_c$, with the scaling slope known from the above. 

As in the previous seascape and demographic noise cases, we can attain the second moment simply by taking the expectation of both sides of the dynamics to get $\Exp{y^2} = (\mu/a)\Exp{y}$.  From here it is easy to see that 
\begin{equation} \label{mix_ratio_sup}
\frac{\Exp{y^2}}{\Exp{y}^2} \propto \frac{1}{\yb} \propto \frac{1}{\delta\mu}.
\end{equation} 
So despite the presence of seascape noise, the scaling of the moment ratio near extinction is identical to the purely demographic noise case.

\end{document}